\documentclass[hidelinks]{ametsocV6.1_ArXiv}
\usepackage{physics}

\title{Recovering Quasi-Biennial Oscillations from Chaos}
\authors{Xavier Chartrand,\aff{a}
Louis-Philippe Nadeau,\aff{a}\correspondingauthor{Xavier Chartrand xavier.chartrand@uqar.ca}
Antoine Venaille,\aff{b}
}
\affiliation{\aff{a}{Institut des Sciences de la Mer de Rimouski, Universit\'e du Qu\'ebec \`a Rimouski, Rimouski, Qu\'ebec G5L 3A1, Canada} \\
\aff{b}{Univ Lyon, ENS de Lyon, Univ Claude Bernard, CNRS, Laboratoire de Physique, F-69342 Lyon, France}
}

\abstract{
The Quasi-Biennial Oscillation (QBO) is understood to result from wave-mean-flow interactions, but the reasons for its relative stability remain a subject of ongoing debate. In addition, consensus has yet to be reached regarding the respective roles of different equatorial wave types in shaping the QBO's characteristics. Here we employ Holton-Lindzen-Plumb's quasilinear model to shed light on the robustness of periodic behavior in the presence of multiple wave forcings. A comprehensive examination of the various dynamical regimes in this model reveals that increased vertical wave propagation at higher altitudes favors periodicity. In the case of single standing wave forcing, enhanced vertical propagation is controlled by the wave attenuation length scale. The occurrence of non-periodic states at high forcing amplitudes is explained by the excitation of high vertical unstable modes. Increasing the attenuation length scale prevents the emergence of such modes. When multiple wave forcing is considered, the mean flow generated by a dominant primary wave facilitates greater vertical propagation of a perturbation wave. Raising the altitude where most of the wave damping occurs favors periodicity by preventing the development of secondary jets responsible for the aperiodic behavior. This mechanism underscores the potential role of internal gravity waves in supporting the periodicity of a QBO primarily driven by planetary waves.
}

\begin{document}
\maketitle

\section{Introduction}\label{sec:introduction}
The Quasi-Biennial Oscillation (QBO) of equatorial winds, once considered a hallmark of low-frequency periodic phenomena emerging from the complex interplay of waves and turbulence \citep{Baldwin2001}, has exhibited a surprising departure from its regularity in recent years. Notably, interruptions and the emergence of a new vertical mode in 2016 and 2020, have raised questions about the long-standing assumption of its steady periodicity \citep[e.g., ][]{Newman2016,Osprey2016}. These disruptions, coupled with similar observations of irregular wind reversals on other planets \citep{Fletcher2017}, have cast uncertainty on the robustness of this equatorial oscillation on Earth \citep{Anstey2022}.

The possibility of non-periodic states in equatorial wind reversals, particularly when wave forcing is intensified, has been documented in various idealized studies. These include simulations based on a 1D quasilinear model \citep{Kim2001,Renaud2019,Leard2020}, 2D direct numerical simulations of stratified fluids subject to lower boundary forcing \citep{Wedi2006,Couston2018,Renaud2019}, and 3D general circulation simulations of gas giants and brown dwarfs \citep{Showman2019}. These findings highlight the absence of a definitive explanation for the remarkable periodicity of the QBO on Earth, prompting the question of why the atmosphere appears to be finely tuned to parameters that correspond to periodic solutions.

In this study, building upon previous work by \cite{Saravanan1990} and \cite{Leard2020}, we delve into the role of multiple wave forcing and critical layers in promoting periodicity. We use the quasilinear model originally developed by Holton, Lindzen, and Plumb \citep{Holton1972,Plumb1977}. This model serves as an invaluable tool for exploring the dynamics of equatorial wind reversals, replicating fundamental aspects of wave--mean-flow interactions, while allowing the exploration of wide regions of the parameter space.

Within the framework of the quasilinear model, several mechanisms have been identified that can support the restoration of periodic states. A first mechanism is due to tropical upwelling, represented as a constant-in-time, bottom-intensified vertical transport term in zonally averaged zonal momentum equations \citep{Saravanan1990}. This term effectively shifts the range of observable periodic states to higher wave forcing values in the parameter space \citep{Rajendran2016}. Upwelling's significance in the altitude range connecting the upper troposphere and the lower stratosphere in Earth's atmosphere is now widely acknowledged \citep{Match2019,Match2021}.

A second mechanism is due to the introduction of a forcing term with prescribed low-frequency oscillations such as a seasonal cycle \citep{Read2012}. Such periodic forcing can result from various factors, including seasonal modulations of upwelling \citep{Rajendran2016}, coupling with semi-annual oscillations in the upper stratosphere \citep{Dunkerton1997}, or seasonal variations of extratropical Rossby wave momentum flux \citep{Bardet2022}.

A third mechanism was introduced by \cite{Leard2020}, suggesting periodicity recovery through the redistribution of forcing associated with a monochromatic wave over a broad spectrum of frequencies via multimodal forcing at the stratosphere's base. They hypothesize that high-frequency waves, which tend to yield more periodic oscillations for a given forcing, become dominant in governing mean-flow oscillations beyond a specific threshold. However, since this forcing redistribution simultaneously reduces the amplitude of the primary wave while increasing the amplitude of high-frequency waves, the role of waves other than the high-frequency ones remains challenging to decipher.

Here, we uncover two additional mechanisms that favor periodicity recovery in the absence of a seasonal cycle and tropical upwelling. Firstly, we emphasize the critical role played by a geometrical parameter that compares the depth of the stratosphere to the typical wave attenuation length scale. We demonstrate that an increase in the attenuation length scale promotes the resurgence of periodic regimes in cases with monochromatic wave forcing. Secondly, we explore the role of multimodal forcing by considering a scenario with two pairs of waves: a fixed background standing wave and a perturbation standing wave, each having distinct wavenumbers and frequencies. Results show substantial regions of parameter space wherein the perturbation drives the system to return to a periodic state, exhibiting a wind structure reminiscent of the Quasi-Biennial Oscillation. Periodicity recovery occurs whether or not waves with higher frequencies govern zonal wind oscillations. Our analysis points to a synchronization mechanism, which coordinates the descent rate of critical layers of both types of waves, preventing the growth of higher vertical modes and explaining the periodicity recovery.

In Section \ref{sec:model}, we introduce the quasilinear model. Section \ref{sec:results} presents the model results, followed by a concluding discussion in Section \ref{sec:conclusion}.

\section{Model}\label{sec:model}
The quasilinear model provides a description of the coupled interaction between internal gravity waves and zonal winds in a stratified fluid \citep{Plumb1977,Vallis2017}. In this model, the stratosphere is represented as a horizontally averaged one-dimensional vertical profile of linearly stratified fluid with a constant buoyancy frequency, denoted as $N$. The model tracks the evolution of zonally averaged zonal winds, $\overline{u}$, under the influence of waves or deviations to the zonal mean, represented as $u'$. Forcing from a wave generator at the lower boundary mimics the impact of the tropopause on stratospheric dynamics. The fluid is confined within a layer of depth $H$, and the mean flow is subject to stress-free and no-flow boundary conditions at the top and bottom, respectively. Additionally, a radiation boundary is applied to the waves at the top boundary.

In this study, our primary focus is on a scenario in which the wave generator emits a maximum of two pairs of symmetric and counterpropagating waves. The first pair, indexed as $b$, is specifically designed to establish the background state of the QBO. Our objective is to investigate how this background state is perturbed by the introduction of another pair, referred to as $p$. The model describes the coupled interaction between the mean flow and the waves as
\begin{align}
\label{eq:momentum_dyn}
\partial_t\overline{u} & = \nu \partial_{zz} \overline{u} - \sum_i \partial_z \overline{u'w'}_i \,, \quad i \in \{b,p\} \,, \\
\label{eq:momentum_flux_wkb}
\overline{u'w'}_i & = F_i \exp\left(-\frac{1}{h_i}\int^z_0\frac{c_i^2}{(\overline{u}-c_i)^2} \, \mathrm{d}z \right) - F_i \exp\left(-\frac{1}{h_i}\int^z_0\frac{c_i^2}{(\overline{u} + c_i)^2} \, \mathrm{d}z \right) \,,
\end{align}
where Eq. \eqref{eq:momentum_dyn} describes the time evolution of the zonal winds forced by the waves' Reynolds stresses, $\overline{u'w'}_i$, and Eq. \eqref{eq:momentum_flux_wkb} describes how these Reynolds stresses are in turn modulated by the zonal winds. Within this framework, the parameters include $\nu$ representing the eddy viscosity, $\pm F_i$ denoting the imposed wave momentum flux for the westward $(-)$ and eastward $(+)$ propagating waves, $\pm c_i$ representing the waves' horizontal phase speeds, and $h_i$ describing the typical vertical attenuation length scale of the wave field. The model is complemented by two boundary conditions\footnote{Our parameterization of wave momentum flux assumes implicitly a radiation condition for waves in the upper stratosphere, such that wave momentum leaks out of the domain. Changing these boundary conditions would change the expression of $\overline{u'w'}$ and details of the phase diagram in the case $\widehat{h} > 1$, but not the mechanisms described in this study.} for the zonal winds:
\begin{equation}
\overline{u} = 0 \,\,\, \text{at} \,\,\, z = 0 \,, \quad \partial_z \overline{u} = 0 \,\,\, \text{at} \,\,\, z = H \,.
\end{equation}
This model can be derived from the nonrotating primitive equations describing the motion of a stratified fluid on the equatorial plane, combined with a quasilinear approach. The derivation consists of three main steps. First, the dynamics are linearized around a prescribed mean flow and the wave field is computed assuming that the mean flow is frozen-in-time. Second, the Reynolds stresses induced by this wave field are computed in the large-time limit, assuming a scale separation between the vertical oscillations of the wave field and the vertical variations of the winds. The final step involves parameterizing the Reynolds stresses using a WKB (Wentzel-Kramers-Brillouin) approximation to evolve the mean flow. Detailed discussions of the underlying assumptions leading to this model can be found in \cite{Renaud2020}.

Equation \eqref{eq:momentum_flux_wkb} for the Reynolds stresses is based on the assumption that wave field dissipation is primarily caused by radiative damping, which occurs at a constant rate denoted as $\gamma$. This parameter determines the attenuation length scale, denoted as $h_i$, for a wave with a given frequency $\omega_i$ and wavenumber $k_i$ through the relationship:
\begin{equation} \label{eq:wave_length}
h_i = \frac{\omega_i^2}{\gamma N k_i} \,.
\end{equation}
For a given wave, fixing the parameter couple $(\omega_i,k_i)$ is equivalent to fixing $(c_i, h_i)$, as the phase speed is given by $c_i = \omega_i/k_i$. This relationship holds under the condition that the buoyancy frequency and damping rate remain steady in time and consistent across the vertical axis. The attenuation length scale, in combination with the forcing amplitude $F_i$ and the phase speed $c_i$, provides an estimate for the characteristic timescale of mean flow reversals for a single monochromatic forcing, which can be expressed as $\tau_i = c_i h_i/F_i$ \citep{Vallis2017}.

We emphasize that the parameter $h_i$ characterizes the typical exponential decay scale of the wave forcing profile in the absence of a mean flow. However, in the presence of a mean flow, the local attenuation length scale at a given altitude is determined by replacing $\omega_i$ with the Doppler-shifted frequency $\omega_i - k_i\overline{u}(z)$ in Equation \eqref{eq:wave_length}. This modification accounts for the influence of the mean flow on the attenuation of the wave field, which in turns modifies vertical profile of the wave forcing. In the case of a constant mean flow, a shallower decay scale occurs when the phase speed $c_i$ shares the same sign as the mean flow $\overline{u}$, and conversely when $c_i$ has the opposite sign to $\overline{u}$. In general, a sheared mean flow leads to significant deviations from an exponential wave forcing profile.

A crucial effect of this Doppler-shift arises when the magnitude of the mean flow $\overline{u}$ becomes comparable to $c_i$. This leads to a situation where the local attenuation length scale approaches zero, effectively causing complete absorption of wave momentum into the mean flow at this particular altitude, thus preventing any wave forcing above it. This phenomenon is referred to as a critical layer. The altitude at which these critical layers can occur has significant implications for the dynamical behavior of the oscillation.

In this study, we explore two distinct approaches to raise this altitude: (i) in the case of monochromatic forcing, by increasing the attenuation length scale $h_i$, and (ii) in the case of wave superposition, by elevating the mean flow $\overline{u}$ to a value greater than the phase speed $c_i$. In both cases, increasing the altitude at which critical layers can occur promotes periodicity.

Note that the critical layer absorption phenomenon takes place irrespective of the magnitude of the damping rate and the specific nature of the dissipation mechanism. While the expression of the attenuation length scale may vary when considering attenuation mechanisms other than radiative damping (such as viscous dissipation or turbulent viscosity accounting for wave-breaking effects), the qualitative nature of the solution is robust.

Equations \eqref{eq:momentum_dyn} and \eqref{eq:momentum_flux_wkb} are solved on a uniform grid with $200$ levels,
using a 3${}^{\mathrm{rd}}$ order Adams-Bashforth time stepping scheme with a timestep $\Delta t = 10^{-4} \tau$, where $\tau$ is the characteristic timescale for mean-flow reversal\footnote{In the case of multi-modal forcing, the timestep is set by the smallest of the two characteristic timescales.}. An implicit diffusion scheme is used for the eddy-viscous term. Injection of momentum for a given wave is tapered above a value of $0.95 c_i$.

\section{Results}\label{sec:results}
\subsection{Periodicity recovery for increasing attenuation length scale}\label{subsec:att_len_scale}
We begin by revisiting the case of a monochromatic standing wave forcing, following the framework initially proposed by \cite{Plumb1977}. In this context, we eliminate perturbations of the background state by setting $\overline{u'w'}_p$ and $F_p$ to zero in Eqs \eqref{eq:momentum_dyn} and \eqref{eq:momentum_flux_wkb}. For simplicity, we temporarily drop the index $b$ of the background state, as we are dealing with a single pair of waves. This context reduces the model's dynamics to two dimensionless control parameters:
\begin{equation}
Re = \frac{F h}{\nu c} \,, \quad \widehat{h} = \frac{h}{H}.
\end{equation}
\begin{figure*}[h!]
\centering
\makebox[\textwidth][c]{\includegraphics[width=44pc]{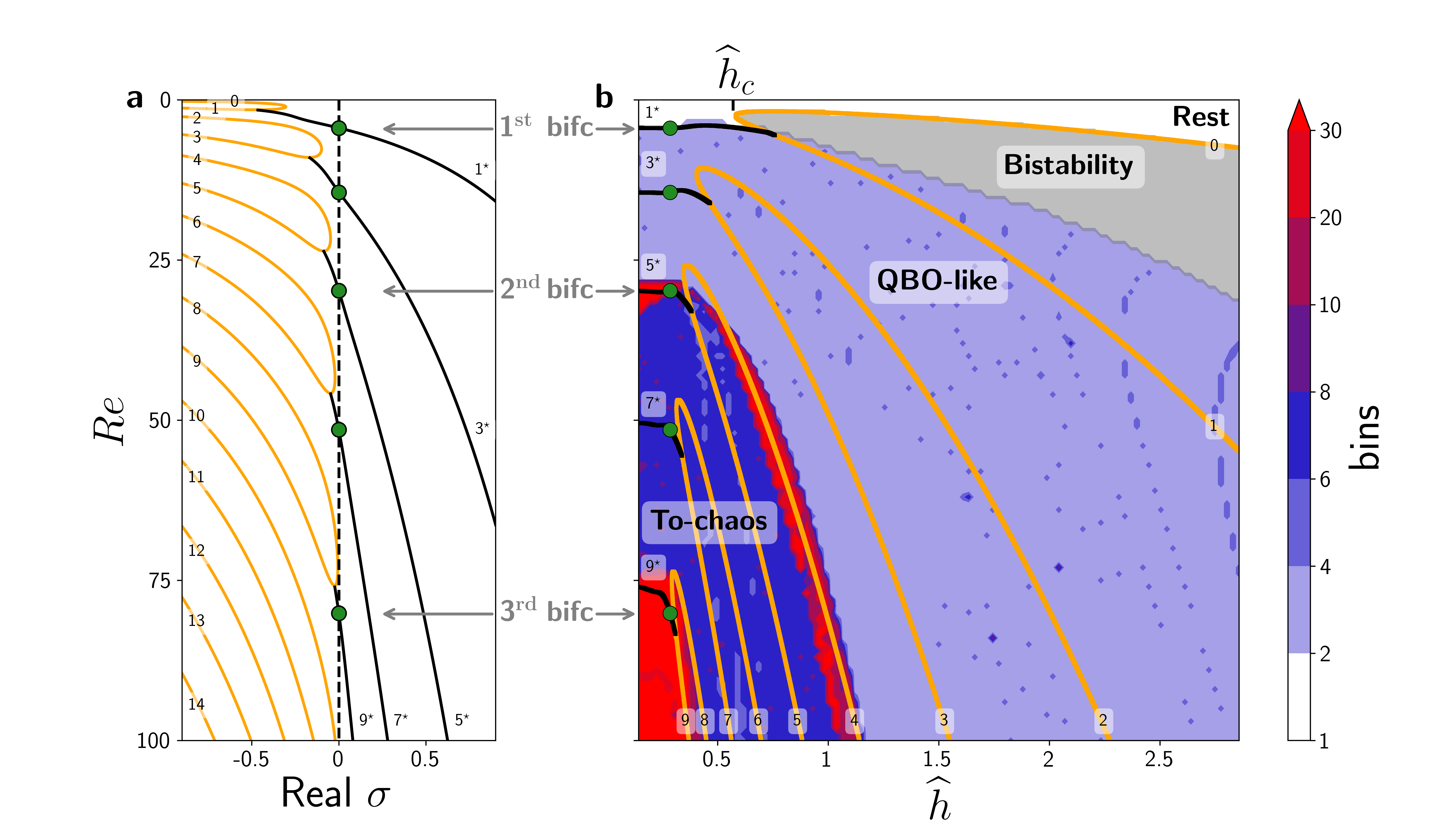}}
\caption{Bifurcation diagram for a monochromatic standing wave forcing. (a) Growth rate of the linearized system for $\widehat{h} = 1/3.5$. Black lines with starred indices: oscillatory modes ($\mathrm{Im}(\sigma) \neq 0$). Orange lines: non-oscillatory modes ($\mathrm{Im}(\sigma) = 0$). Indices refer to the vertical structure of the modes. (b) Dynamical regimes of the quasilinear model in $(\widehat{h}, Re)$ space, using Poincar\'e map metric. Black lines indicate the appearance of an unstable oscillatory mode in the linearized system, orange lines correspond to non-oscillatory ones. Green dots correspond to those in the panel (a).}
\label{fig:1Dsimu_ReHspace}
\end{figure*}
Figure~\ref{fig:1Dsimu_ReHspace}b presents the bifurcation diagram obtained by varying of both dimensionless control parameters. Bifurcation diagrams are constructed using Poincaré maps, where values of $\overline{u}$ near the top of the domain that intersect with $\overline{u} = 0$ near the bottom are grouped into bins. The shading in the Figure corresponds to the total number of bins populated in a given experiment. Experiments with only one bin indicate a stable resting state, those with two bins represent a QBO-like periodic oscillation, while more than two bins signify the progression toward aperiodicity and chaotic behavior \citep{Renaud2019}. The transition from a resting state to a QBO-like oscillation is referred to as the first bifurcation, whereas the transition from a QBO-like oscillation to higher oscillatory modes is termed the second bifurcation.

In standard textbook treatments, the assumption is often made of a semi-infinite stratosphere, where $\widehat{h} \rightarrow 0$, corresponding to the left of Fig.~\ref{fig:1Dsimu_ReHspace}b. Remarkably, for any fixed value of the forcing parameter $Re$, the system transitions towards more periodic states as $\widehat{h}$ becomes sufficiently large. This implies that, for a finite domain depth, $H$, modulations of wind reversal regimes can be achieved solely by varying the attenuation length scale $h$, which is determined by the wave properties originating in the upper troposphere, regardless of their forcing amplitude.

Focusing on the structure of bifurcation diagrams and considering the entire range of $Re$ as $\widehat{h}$ increases, a new dynamical regime emerges, which is not present in the semi-infinite assumption ($\widehat{h} \rightarrow 0$). This regime is characterized by bistability, represented by the light grey shading in Figure~\ref{fig:1Dsimu_ReHspace}b. Within this bistable regime, two steady non-zero zonal wind solutions coexist, effectively breaking the east-west system symmetry imposed by the wave generator. The introduction of this new dynamical regime has a significant impact on the behavior of the first bifurcation, which occurs between resting states and QBO-like oscillations. This transition is marked by a shift from a supercritical Hopf bifurcation to a subcritical one as $\widehat{h}$ surpasses a critical threshold, denoted as $\widehat{h}_c$. For values of $\widehat{h}$ below $\widehat{h}_c$, the system undergoes a first bifurcation from a state of rest to a QBO-like oscillation. However, when $\widehat{h}$ exceeds $\widehat{h}_c$, the system first transitions to a steady non-zero zonal wind profile before generating QBO-like regimes. The system exhibits hysteresis behavior as $Re$ is varied continuously across this regime. Similar transitions from supercritical to subcritical Hopf bifurcations have been previously reported by \cite{Semin2018} when varying other parameters of the quasilinear model, such as the strength of an additional linear drag, see also \cite{Yoden1988} in the case of asymmetric wave forcing.

\subsubsection{Predicting bifurcations with the linearized system}\label{subsubsec:linearized_sys}
To gain insight into the dynamics explaining the bifurcation diagram of Fig.~\ref{fig:1Dsimu_ReHspace}b, we linearize the model around a state of rest, following the approach of \cite{Semin2018}. Using nondimensional parameters:
\begin{equation}
U = \frac{\overline{u}}{c} \,, \quad Z = \frac{z}{H} \,, \quad T = \frac{t F}{c h} \,,
\end{equation}
and assuming $U \ll 1$, we obtain, at the lowest order, the rescaled linear system:
\begin{equation} \label{eq:linear_u}
\partial_T U = \frac{\widehat{h}^2}{Re} \partial_{ZZ} U + 4 \mathrm{e}^{-Z/\widehat{h}} \left(U - \frac{1}{\widehat{h}}\int_0^Z U \, \mathrm{d}Z' \right) \,,
\end{equation}
with boundary conditions
\begin{equation}
U = 0 \,\,\, \text{at} \,\,\, Z = 0 \,, \quad \partial_Z U = 0 \,\,\, \text{at} \,\,\, Z = 1 \,.
\end{equation}
This system can be turned into an eigenvalue problem, $\sigma \widehat{U} = L \hat{U}$, by writing $U(z,t) = \widehat{U} \mathrm{e}^{\sigma t}$. The eigenspectrum is spanned by solving numerically the integro-differential operator $L$, corresponding to the right-hand side of Eq. \eqref{eq:linear_u}. We first focus on the classical scenario of a semi-infinite stratosphere, characterized by $\widehat{h} \ll 1$, for which the real part of the eigenvalue $\sigma$ is plotted in Fig.~\ref{fig:1Dsimu_ReHspace}a as a function of $Re$. In this Figure, black signifies that a mode is oscillatory ($\mathrm{Im}(\sigma) \neq 0$), while orange signifies a mode that is not oscillatory ($\mathrm{Im}(\sigma) = 0$). Vertical modes are numbered based on the maximum number of zeros or nodes, excluding boundary conditions (see Supplemental Materials). The point at which an eigenvalue changes sign marks the onset of unstable exponential growth (green dots in Fig.~\ref{fig:1Dsimu_ReHspace}a). Within this semi-infinite domain context, the threshold of the first bifurcation in Fig.~\ref{fig:1Dsimu_ReHspace}b corresponds to the instability of the first oscillating mode, denoted as 1${}^\star$. Remarkably, the threshold of the second and third bifurcations, as predicted in the quasilinear simulations, closely aligns with the instability of the third and fifth oscillating mode, referred to as 5${}^\star$ and 9${}^\star$. We conclude that the transition to chaos corresponds to the excitation of unstable modes with an increasing number of alternating horizontal jets in the vertical direction as the Reynolds number is increased.

The onset of instability of each modes is illustrated as black and orange lines in the bifurcation diagram of Figure~\ref{fig:1Dsimu_ReHspace}b. As $\widehat{h}$ increases, all unstable eigenvalues become purely real, as the wave operator $L$ progressively becomes proportional to a Laplacian. This is why any black branch (oscillatory modes) split into two orange branches (non-oscillatory) modes as $\widehat{h}$ is increased in \ref{fig:1Dsimu_ReHspace}b. Consequently, the first QBO bifurcation at large $\widehat{h}$ is different than in the semi-infinite case ($\widehat{h} \ll 1$). For values of $\widehat{h}$ larger than a critical threshold denoted $\widehat{h}_c$, the emergence of the first unstable eigenvalue corresponds to a pitchfork bifurcation of fixed points (from a state of rest to bistability). The threshold of this first bifurcation corresponds to the instability of the first stationary mode, denoted as 0 in Fig.~\ref{fig:1Dsimu_ReHspace}b. This bistable configuration becomes unstable at higher $Re$, with the emergence of a limit cycle around the two fixed points. The transition occurs before the second stationary unstable eigenvalue and is not captured by the linearized dynamics around a state of rest, since this would instead require a linearization around the fixed points. Surprisingly, the threshold of the second bifurcation, originally captured by the instability of the oscillatory-mode-5${}^{\star}$ at $\widehat{h} \ll 1$, coincides with the instability of the stationary-mode-4 as $\widehat{h}$ increases.

It is intriguing that the instability of the eigenmodes of the linearized system is capable of capturing the emergence of successive bifurcations beyond the first one given that they are computed around a state of rest. Additionally, the connection between the instability of certain modes and their impact on bifurcations, while others have limited effect, is also somewhat puzzling. To fully understand these selection rules and explain the predictive ability of the linearized dynamics around a state of rest for subsequent bifurcations, a comprehensive Floquet analysis would be required. However, conducting such an analysis falls beyond the scope of this paper.

Dimensional analysis helps explain why increasing $\widehat{h}$ promotes periodicity. In a semi-infinite stratosphere where $\widehat{h} \rightarrow 0$, $Re$ is the sole dimensionless parameter governing the system's dynamics \cite{Vallis2017}. When $\widehat{h}$ is large, a direct inspection of the equation (\ref{eq:linear_u}) shows that the appearance of new unstable eigenmodes is controlled by $Re/\widehat{h}^2$. This amounts to defining a new Reynolds number based on the domain height $H$ rather than on the attenuation length scale $h$ and on the reduced forcing $F/\widehat{h}$ rather than $F$. In the limit $\widehat{h} \gg 1$, bifurcations occur at finite values of $Re/\widehat{h}^2$, which causes the range of $Re$ between each bifurcation to widen as $\widehat{h}$ increases, as observed in Figure~\ref{fig:1Dsimu_ReHspace}b.

We have seen that the emergence of aperiodic states involves the excitation of higher vertical modes with multiple jets. During their nonlinear evolution, some of these jets grow until reaching the upper-bound set by the critical layers $\overline{u} = c$. In the case of a semi-infinite domain, the altitude at which these critical layers form scales roughly with $h$. As $\widehat{h}$ becomes comparable to $1$ or larger, this altitude scales with the stratosphere height $H$. More generally, for a given $Re$, increasing $h$ implies reduced wave damping, higher altitude wave propagation, and increased vertical homogeneity, concomitantly with periodicity recovery. We hypothesize that this vertical homogeneity in wave forcing promotes periodicity as it hinders the development of higher vertical modes. This hypothesis will be further substantiated in the case of multimodal forcing in Section \ref{sec:results}\ref{subsec:2waves}.

\subsection{A new metric to characterize dynamical regimes using vertical modes}\label{subsec:metric}
\begin{figure*}[h!]
\centering
\includegraphics[width=39pc]{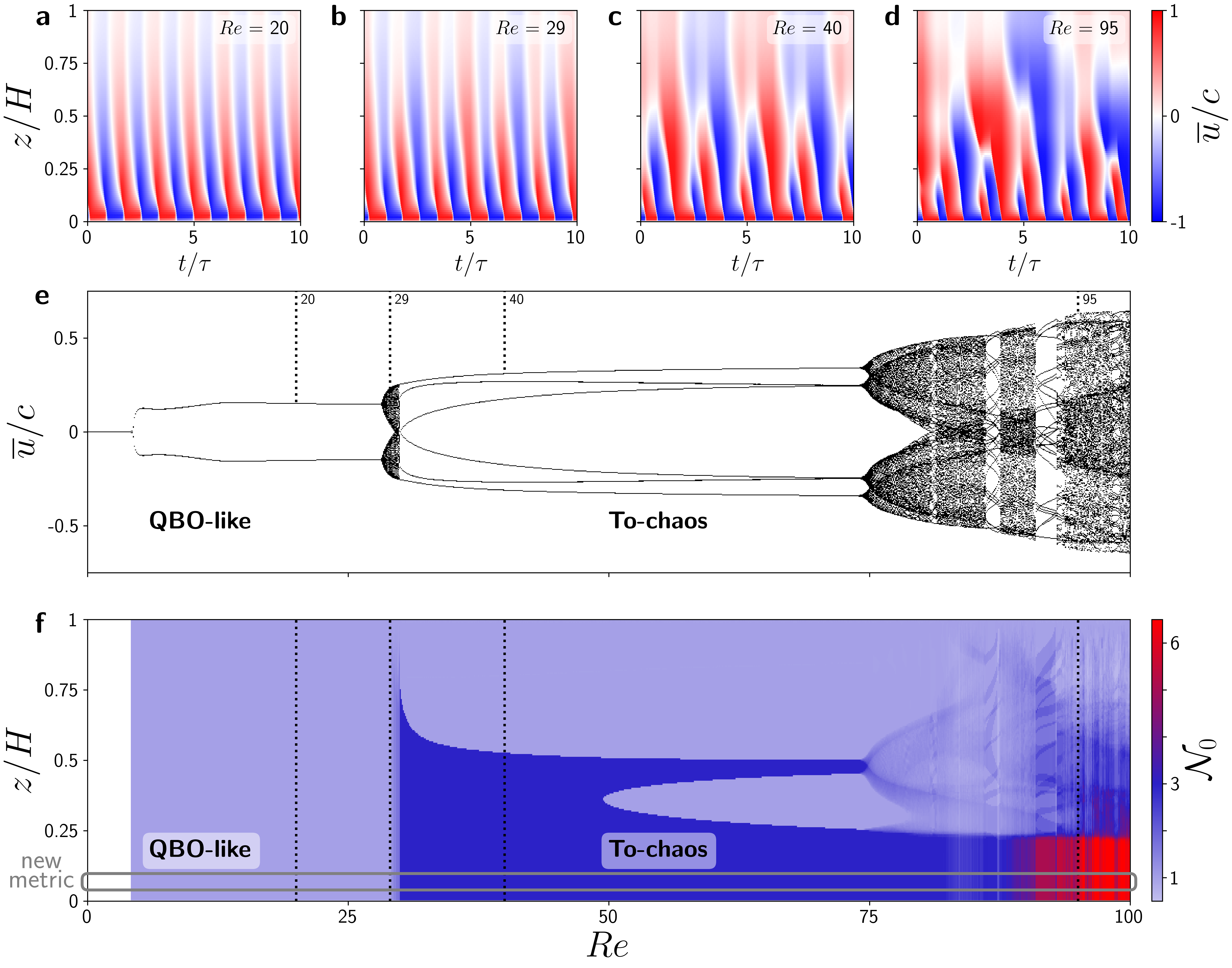}
\caption{(a)-(d) Hovmöller diagrams for the wind speed profile at different Reynolds numbers, for $\widehat{h} = 1/3.5$. (e) Corresponding bifurcation diagram computed with Poincar\'e metric. (f) Same bifurcation diagram computed with the new metric $\mathcal{N}_0$, defined as the ratio between the number of flow reversals on every level to that of the upper one.}
\label{fig:NewMetric}
\end{figure*}

As noted above, the onset of aperiodicity is inherently associated with the instability of higher vertical modes of the linearized system, with the number of alternating jets increasing as the forcing parameter grows. This excitation of higher vertical modes with increasing forcing is manifested in the quasilinear model, as illustrated using Hovmöller plots of the zonal winds in Fig.~\ref{fig:NewMetric}a-d. Periodic oscillations are characterized by a maximum of one downward propagating flow reversal in the vertical at any instant (or 1-node)\footnote{Boundary conditions are excluded from this definition.}. In contrast, strongly forced aperiodic regimes exhibit time periods with multiple downward propagating nodes. Figure~\ref{fig:NewMetric}e depicts a bifurcation diagram of the histograms employing Poincaré maps, for which the information is condensed as a number of bins in Figure~\ref{fig:1Dsimu_ReHspace}b. This is contrasted with Fig.~\ref{fig:NewMetric}f, where the ratio of the number of flow reversals compared to that of the topmost grid point is plotted as a function of altitude. As the forcing is increased, faster flow reversals are observed at the bottom of the stratosphere compared to the top. These higher vertical structures can be interpreted as an entanglement of multiple oscillators with increasing frequency toward the ground (see Fig.~\ref{fig:NewMetric}d). Importantly, each major transition in the bifurcation diagram of Fig.~\ref{fig:NewMetric}e is closely linked to a regime transition in the vertical modes in Fig.~\ref{fig:NewMetric}f.
Based on this observation, we propose a new metric for diagnosing QBO-like regimes that focuses on the vertical structure of the oscillations, rather than employing histograms of Poincaré maps \citep[e.g.,][]{Renaud2019,Leard2020}. For a given total simulated time, this new metric computes the ratio of the number of zeros counted in the bottom boundary layer (grey region in Fig.~\ref{fig:NewMetric}f) compared to that counted near the top. This ratio is loosely related to a vertical mode as it yields a time-averaged number of nodes in the column, excluding periods without zonal wind reversal from this average. The numbering of each vertical mode, explained in Supplemental Materials (Fig.~S2), is based on the maximum number of zeros in the column, excluding boundary conditions. For oscillatory modes, we consider the maximum through a complete oscillation.

Hereafter, states with a ratio of one, meaning a single node or mode-1, will be referred to as QBO-like states. Notably, this metric is better suited to describe the actual Earth QBO, which exhibits fluctuations around its average period while remaining trapped in a mode-1 state prior to 2016. Yet, under this new metric, both periodic and quasiperiodic solutions with mode-1 vertical structure are considered QBO-like. This results in a slight shift in the value of the second bifurcation compared to the Poincaré metric, corresponding to the quasiperiodic region (Fig.~\ref{fig:NewMetric}e,f).

This minor limitation is outweighed by a major advantage compared to Poincaré sections: it exhibits a monotonic increase with both $Re$ and $h$. For example, quasiperiodic regimes are often wedged between periodic and frequency-locked regimes when increasing $Re$ (Fig.~\ref{fig:NewMetric}e). This is diagnosed as a local maximum using Poincaré sections, whereas a monotonic increase is observed using vertical modes. The monotonicity of the metric greatly helps in interpreting the response of the system to a given change in parameter. In particular, if increasing a parameter reduces the number of nodes, it can be interpreted as favoring periodicity. Conversely, if the number of nodes increases with an increasing parameter, it suggests a tendency towards aperiodicity or more complex dynamics. This idea will greatly facilitate the interpretation of the results in the following section, which involve perturbations to the background QBO.
\subsection{Periodicity recovery by perturbations to an aperiodic background state}\label{subsec:2waves}
We now shift the focus to the case of a superposition of multiple counterpropagating waves, which is arguably more realistic than monochromatic wave forcing. As discussed above in the Model section, we consider, for simplicity, the case of only two pairs of counterpropagating waves introduced in Equations \eqref{eq:momentum_dyn} and \eqref{eq:momentum_flux_wkb}. Adding a pair of counterpropagating waves with amplitude $F_p > 0$, frequency $\omega_p$ and wavenumber $k_p$ implies three new nondimensional parameters:
\begin{equation}
\frac{F_p}{F_b} \,, \quad \frac{c_p}{c_b} \,, \quad \frac{h_p}{h_b}.
\end{equation}
Having already discussed in depth the role of the attenuation length scale in the monochromatic case, we now constrain ourselves to the case $h_p = h_b \ll H$ in the following discussion. This restriction does not compromise the generality of the proposed mechanism for periodicity restoration. Below, we describe how the introduction of a second wave exerts a stabilizing influence on the system by preventing the excitation of higher vertical modes responsible for aperiodicity.

\begin{figure*}[h!]
\centering
\includegraphics[width=30pc]{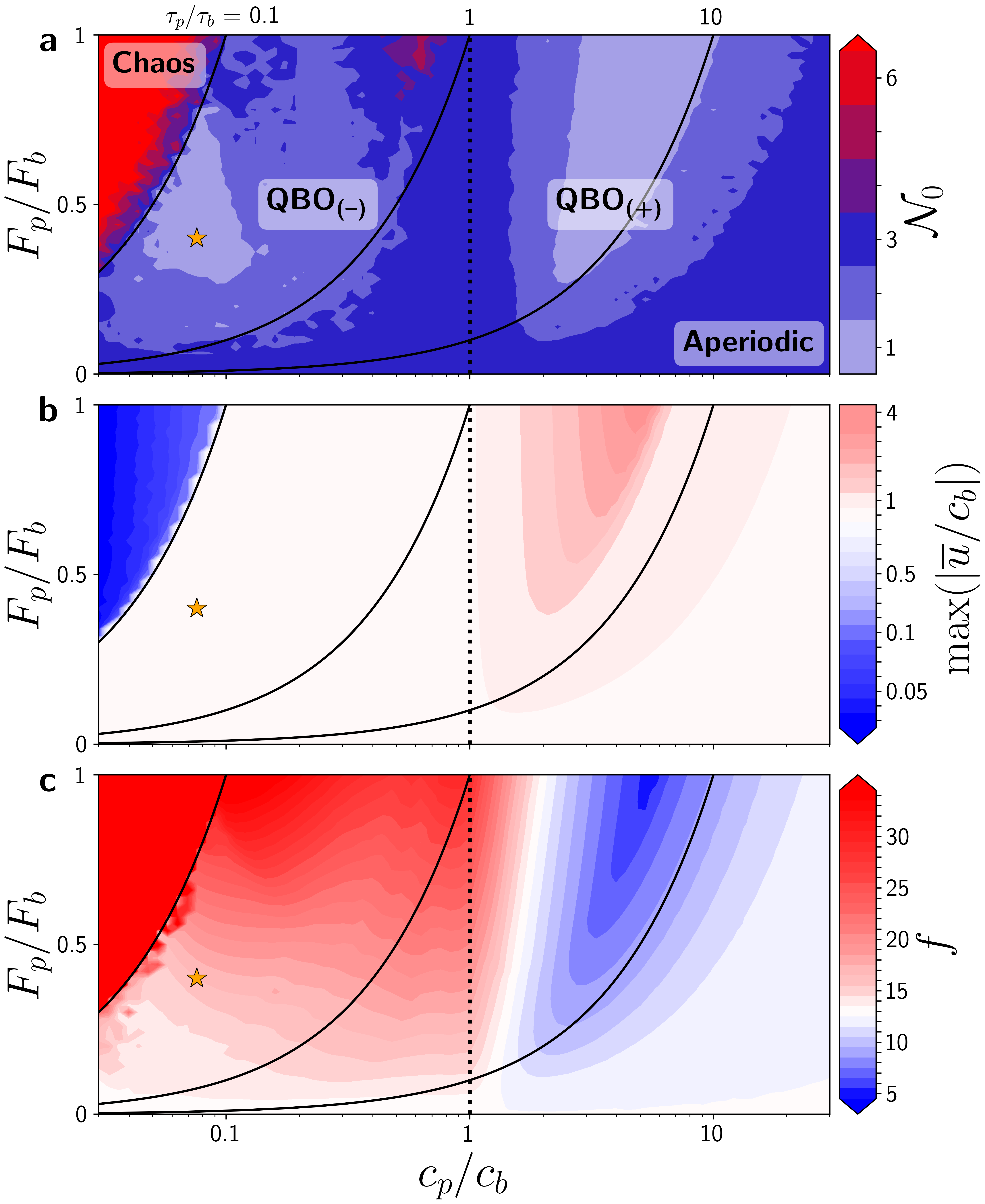}
\caption{(a) Bifurcation diagram for the case of multiple wave forcing. See text and Figure~\ref{fig:NewMetric} for a definition of the metric $\mathcal{N}_0$. (b) Maximal wind speed recorded in each regime. (c) Frequency of wind reversals computed from a time series of the fastest oscillations near the lower boundary in each regime. In each panel, thick black contours indicate isolines of $\tau_p/\tau_b$.}
\label{fig:1Dsimu_2waves}
\end{figure*}

In order to disentangle the role of the perturbation from that of the base state, we set $F_b$ to a fixed value characterized by a mode-3 oscillation, beyond the second bifurcation, which corresponds to the solution shown at Fig.~\ref{fig:NewMetric}c. We then conduct a series of experiments where we incrementally increase the perturbation $F_p$ while keeping $F_b$ constant --- thus increasing the total forcing --- for a wide range of phase speed ratios $c_p/c_b$. One might intuitively anticipate that as the total forcing increases, the system would transition towards higher vertical modes and chaos, as noted by \cite{Renaud2019}. Paradoxically, our results, depicted in Fig.~\ref{fig:1Dsimu_2waves}a, reveal a parameter space featuring instead two prominent regions where the flow transitions back to a periodic QBO-like oscillation: one for $c_p < c_b$, denoted as QBO$_{(-)}$, and another for $c_p > c_b$, denoted as QBO$_{(+)}$. Apart from a small region in the upper left corner of Fig.~\ref{fig:1Dsimu_2waves}a ($c_p \ll c_b$), where higher vertical modes with chaotic reversals are excited, no significant changes are observed elsewhere.
To understand the primary drivers in these regions, we consider the maximum velocity of the mean flow, displayed in Fig.~\ref{fig:1Dsimu_2waves}b. When the background wave with phase speed $c_b$ governs the system, the maximum speed of the mean flow is $ \overline{u}_{max} \approx c_b $. In most of the parameter space, regimes are dominated by the background wave. There are two notable exceptions. Firstly, the system undergoes an abrupt transition when $c_p \ll c_b$, where the mean flow decelerates suddenly (Fig.~\ref{fig:1Dsimu_2waves}b) as it becomes governed by the perturbation associated with chaotic high vertical modes (Fig.~\ref{fig:1Dsimu_2waves}a). Secondly, a smooth transition occurs for $c_p > c_b$, where the mean flow gradually accelerates as it increasingly falls under the influence of the perturbation. This smooth transition for $c_p > c_b$ aligns with a decrease in the frequency of wind reversals (as indicated in Fig.~\ref{fig:1Dsimu_2waves}c) and is closely linked to the QBO$_{(+)}$ region of periodicity recovery observed in Fig.~\ref{fig:1Dsimu_2waves}a. This contrasts with the region QBO$_{(-)}$, where the frequency of wind reversals increases while no change in mean flow is observed. Below, we focus individually on each of the periodic regions.

In the QBO$_{(-)}$ region, where the background wave governs the system, two types of critical levels can emerge: (i) critical levels associated with the background waves, which occur when the maximum wind speed approaches the phase speed of the background waves ($\overline{u}_{\max} \approx \pm c_b$); and (ii) critical levels linked to the perturbation waves, which occur when the mean wind speed matches the horizontal phase speed of the perturbation waves ($\overline{u} = \pm c_p$).

\begin{figure*}[h!]
\centering
\includegraphics[width=30pc]{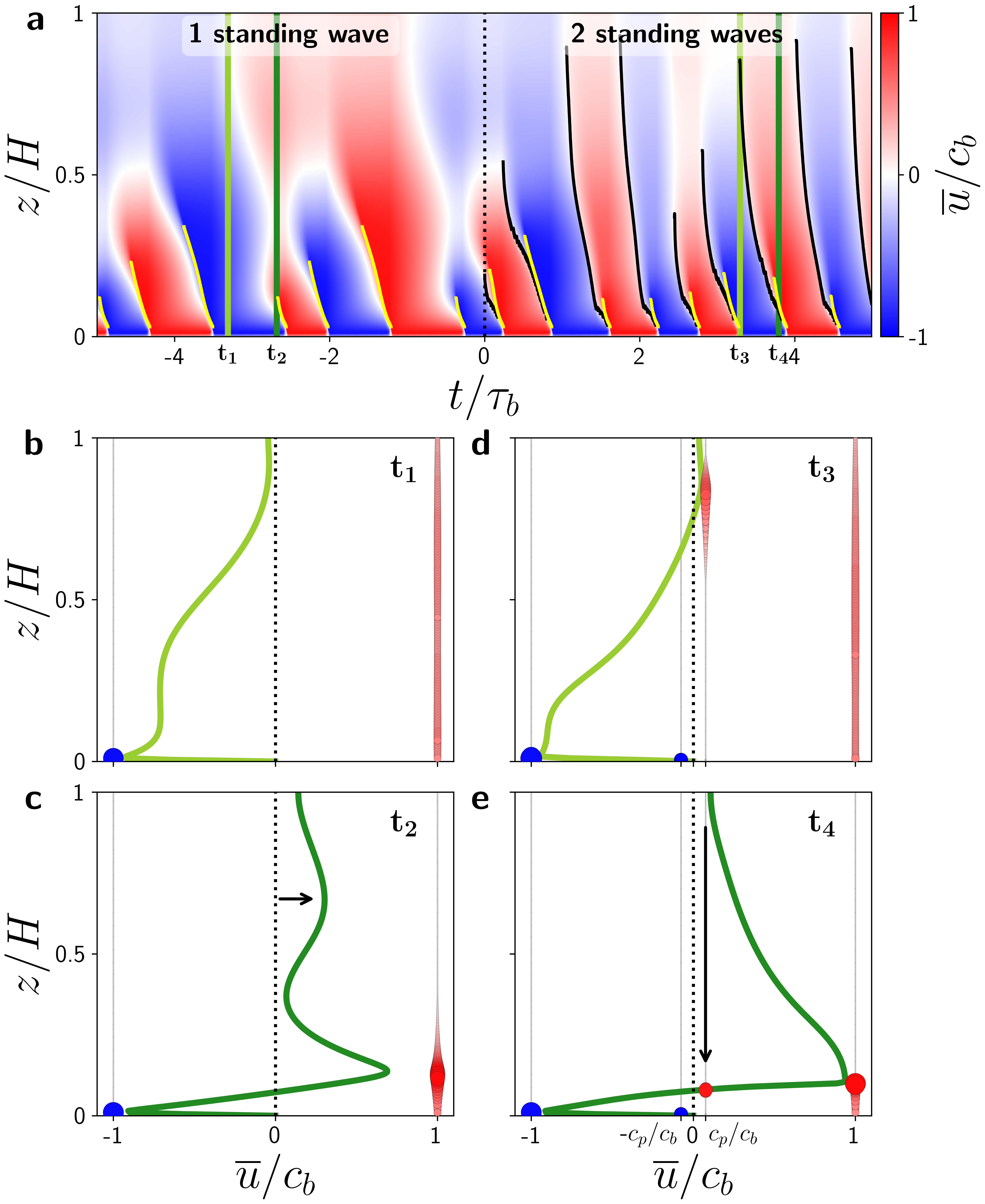}
\caption{Transition from mode-3 to QBO-like oscillations with the addition of a perturbation wave at $t = 0$, corresponding to the orange star in Figure~\ref{fig:1Dsimu_2waves}. (a) Hovmöller diagram. Black and yellow contours trace the downward propagation of critical layers of the perturbation and the background waves, respectively. (b) and (c) Instantaneous vertical profiles of the mean flow (green line) under monochromatic wave forcing, with blue and red dots representing westward and eastward Reynolds stress derivatives, respectively (dot size is proportional to amplitude). (d) and (e) Same for multiple wave forcing, where Reynolds stresses are decomposed into background wave (at $\overline{u} = \pm c_b$) and perturbation wave (at $\overline{u} = \pm c_p$) contributions.}
\label{fig:CL_2waves}
\end{figure*}

To gain insight into the role of each type of critical level, we conduct an experiment depicted in Fig.~\ref{fig:CL_2waves}, where a perturbation is introduced at $t = 0$ from the reference experiment using monochromatic forcing.
After just a few oscillation cycles following the introduction of the perturbation wave, a QBO-like cycle is recovered (Fig.~\ref{fig:CL_2waves}a), corresponding to the experiment marked with an orange star in the Figure~\ref{fig:1Dsimu_2waves}. In both the monochromatic and multimodal cases, we analyze two snapshots typical of the westward phase of the mean winds ($\overline{u} < 0$), as well as their associated wave forcing. The first snapshot is taken soon after the previous wind reversal, while the second snapshot is right before the next wind reversal (as shown in Fig.~\ref{fig:CL_2waves}).

In both cases, the forcing from the background waves is similar at both times (Fig.~\ref{fig:CL_2waves}b-e). However, in the multimodal case, an additional eastward forcing from the perturbation is initially introduced in the upper part of the stratosphere (see $t_3$ in Fig.~\ref{fig:CL_2waves}e). This high-altitude eastward forcing is a consequence of the Doppler shift's impact on the local attenuation length scale\footnote{The Doppler-shifted attenuation length scale of the perturbation in the presence of a mean flow, where $\overline{u}\sim c_b$ is much larger than the phase speed $c_p \ll c_b$, scales as $(c_b/c_p)^2 h$ (as explained in Section \ref{sec:model}).}, which allows the perturbation wave with an eastward phase speed to propagate upward with minimal attenuation until it reaches the upper part of the stratosphere, where the winds are much weaker. This efficient transport leads to high-altitude deposition of eastward momentum from the perturbation and to the development of a fast downward propagating critical layer that closely follows the zero-wind line (Fig.~\ref{fig:CL_2waves}d,e). This fast-moving critical layer prevents the excitation of secondary westward jets or higher vertical modes responsible for aperiodicity, by smoothing the vertical wind profile (compare Fig.~\ref{fig:CL_2waves}c,e).

Periodicity recovery results from a synchronization mechanism between the fast descent of the perturbation-driven zero-wind line and the background-driven mean-wind oscillation. Synchronization occurs when the perturbation-driven critical layers are sufficiently faster than the background-driven critical layers to prevent the growth of secondary fronts forced by the background wave. The typical time scale associated with the descent of background-driven critical layers is $\tau_b = c_b h_b/F_b$, while that of the perturbation-driven critical layers is $\tau_p = c_p h_p/F_p$, assuming that all the momentum of the perturbation wave is efficiently used to move the location where $\overline{u} = c_p$ downward. A periodic state is thus expected when $\tau_p < \tau_b$.

In each panel of Fig.~\ref{fig:1Dsimu_2waves}, lines representing the ratio $\tau_p/\tau_b = 0.1, 1, 10$ are displayed. The region QBO$_{(-)}$ is well represented by the area between $\tau_p/\tau_b = 0.1$ and $1$, as expected from the above analysis. In this region a periodic behavior with a QBO-like vertical structure emerges from the combination of perturbed and background wave forcing, which would independently lead to nonperiodic behavior. This synchronization mechanism ceases to be effective when the critical levels associated with the perturbation are fast enough to govern the oscillation itself, which becomes chaotic at $\tau_p/\tau_b \approx 0.1$.

The occurrence of the region QBO$_{(+)}$ coinciding with the line of $\tau_p/\tau_b = 10$ suggests that a similar synchronization mechanism is at play in this second regime of periodicity recovery. However, in this case, the roles of the background and perturbation are somewhat reversed: critical levels associated with the background forcing occur progressively at higher altitudes as the maximum mean flow gradually accelerates when the oscillation falls under the influence of the perturbation (see Fig.~S3). Additionally, the transfer of momentum from the perturbation to the background state is much less efficient than in QBO$_{(-)}$ region due to a compensation effect between the wave momentum flux induced by the eastward-moving and westward-moving perturbations, which prevents these perturbations from reaching critical levels. An extreme case is observed when $c_p / c_b \gg 1$, where the perturbation has virtually no effect on the wind reversals. This is evident in Fig.~\ref{fig:1Dsimu_2waves}, where QBO properties remain unchanged beyond $c_p = 10 c_b$. This occurs because the wave momentum flux induced by the eastward-moving perturbation becomes nearly equal to the contribution from the westward-moving perturbation when $\overline{u}/c_p \ll 1$, as described by Eq. \eqref{eq:momentum_flux_wkb}. For finite values of $c_p/c_b > 1$, the primary effect of the perturbation is to shift the maxima of the wind profile to values greater than the critical level $c_b$. Assuming that most of the descent of these maxima remains dominated by contributions from the background wave forcing, the period of the QBO can be estimated as $\overline{u}_{\mathrm{max}} h_b/F_b$, with $\overline{u}_{\mathrm{max}}$ being an increasing function of $F_p$. This explains the decrease in frequency associated with the QBO$_{(+)}$ region in Fig.~\ref{fig:1Dsimu_2waves}c.

In summary, perturbation waves with phase speeds $c_p < c_b$ lead to a faster descent of the QBO front to the bottom of the stratosphere, resulting in a shorter period of wind reversals. On the other hand, waves with $c_p > c_b$ amplify the amplitude of the QBO front, leading to an extended period of reversals. In both cases, the presence of a wave with a slower phase speed prevents the growth of unstable modes of the larger wave due to resonance between two timescales: any local extremum in the velocity profile is shifted downward before its amplitude reaches a critical level. This effectively prevents the excitation of higher vertical modes responsible for aperiodicity and explains the observed periodicity recovery regions characterized by a vertical mode-1 akin to the QBO. This phenomenon is all the more surprising as it emerges from the combination of perturbed and background wave forcing that independently would lead to nonperiodic reversals.

\section{Conclusion}\label{sec:conclusion}
Building upon the insights gained from the classical monochromatic scenario, we have proposed a physical mechanism for intrinsic synchronization in internal-wave--mean-flow interactions with multiple-wave forcing. The key concept is that raising the altitude at which wave momentum is deposited promotes periodic behavior. This can be achieved by modifying the attenuation length scale in the monochromatic case or through multimodal wave forcing, where the mean flow generated by the faster wave significantly alters the Doppler-shifted attenuation length scale of the slower wave. To maintain simplicity, we focused on a scenario involving only two pairs of counterpropagating waves, which lays the groundwork for interpreting more complex configurations involving a wider distribution of waves generated in the upper troposphere.

Using a quasilinear model similar to ours, forced with a Gaussian frequency spectrum, \cite{Leard2020} found that increasing the variance of the distribution, while maintaining a constant total forcing, favors periodic behavior.
The resulting QBO-like periodic regime exhibited significantly higher velocities compared to the chaotic monochromatic forcing, leading the authors to suggest that high-frequency waves, given a fixed horizontal wave number, played a crucial role in shaping QBO characteristics. Translating their observation into our specific framework, we associate this recovery of periodicity to the region denoted as QBO${}_{(+)}$, for which a similar zonal winds amplification is observed. Importantly, our study unveiled another region of periodicity recovery, QBO${}_{(-)}$, where the phase speed of the perturbation wave is smaller than the background wave phase speed. In contrast to the periodicity recovery in the QBO${}_{(+)}$ region, the maximum zonal wind remains unchanged in QBO${}_{(-)}$ since the mean flow is governed by the background wave in this region. Thus, beyond modulations in the forcing frequency, our study highlights the critical influence of perturbation phase speed variations, achieved by simultaneous adjustments in frequency and wavenumber.

In the context of a more realistic wave representation, the respective roles of the different classes of equatorial waves remains a subject of debate \citep[e.g., ][]{Pahlavan2021,Holt2022}. While planetary-scale waves are widely acknowledged as a central driver of the oscillation, previous numerical studies of the atmospheric QBO have emphasized the need to parameterize the effects of small-scale internal gravity waves in order to induce a QBO-like regime \citep{Giorgetta2006,Lott2013}.

In the framework of our study, the primary standing wave serve as a rudimentary representation of the influence exerted by planetary Rossby, Yanai, and Kelvin waves, while the perturbation standing wave can be understood as the result of smaller-scale and higher-frequency internal waves generated by localized convection events.

Our study demonstrates the crucial role of the zonal phase speed ratio, denoted as $c_p/c_b$, which characterizes an important relationship between background and perturbation waves. The phase speed of convectively coupled equatorial Kelvin waves, estimated at about $15$ m/s, serves as a reference value for $c_b$. These waves exhibit a vertical wavelength of $\pi/H_t$, where $H_t$ represents the tropospheric height based on the equatorial Rossby radius of deformation, $\sqrt{c_b/\beta}$. We assume that perturbations arise from localized convection events that generate internal gravity waves with similar vertical length scales, namely $\pi/H_t$. However, their horizontal length scales are significantly smaller than the equatorial radius of deformation. This results in a perturbation phase speed with an absolute value close to $c_b$. Nonetheless, the projection of this phase speed in the zonal direction is on average smaller than $c_b$ in the case of oblique ray propagation with respect to the equator \citep{Kim2023}. Synchronization induced by such oblique waves would then correspond to the region QBO$_{(-)}$ highlighted in our study.

To establish this interpretation on firmer ground, we will need to generalize our study to a 3D configuration on the stratified equatorial beta plane \citep{Plumb1982}. In this expanded model, the attenuation length scale for Rossby and Yanai waves will differ from the scales used in our study, which concentrated solely on internal gravity waves. It should be reminded that we also neglected additional effects such as coupling with an external low frequency oscillator \citep{Read2012}, stochasticity \citep{Saravanan1990,Wedi2006}, wave-wave interactions \citep{Couston2018}, transient behavior of the wave field \citep{Dunkerton1981}. These effects are currently under investigation for efficient parameterizations of unresolved internal gravity waves in climate model \citep[e.g.,][]{Achatz2023}. Ensuring the robustness of the periodicity recovery mechanism described in our study will require thorough testing against these influences.

\acknowledgments
We acknowledge Cerasela Calugaru and the whole team from PSMN/CBP at ENS de Lyon for computing facilities. We warmly thank Nicolas Perez for his valuable help in performing the linear analysis with the Dedalus solver. We also thank Louis Couston and Jason Reneuve for useful discussions on wind reversals with multiple wave forcing. Louis-Philippe Nadeau is partly supported
through NSERC Award RGPIN-2022-04306.

\datastatement
The model used for this study is available on GitHub at https://github.com/xavier-chartrand/qbo1d.git. Model outputs are archived on PSMN clusters and remain available upon request from XC.

\bibliographystyle{ametsocV6}
\bibliography{references}

\clearpage
\setcounter{figure}{0}
\setcounter{equation}{0}
\renewcommand{\figurename}{Fig. S}
\renewcommand{\theequation}{S\arabic{equation}}
\onecolumn
\begin{center}
\large\bf Supplemental Materials to: Recovering Quasi-Biennial Oscillations from Chaos
\end{center}
\section*{\centering I. Computation of the eigenvalue problem}
In the following, we present the approach to solve the eigenvalue problem arising from the Holton-Lindzen-Plumb model when linearized around a state of rest, along with a detailed picture of eigenvalues $\sigma$ and eigenmodes $\widehat{U}$. Recalling our choice of dimensionless parameters, namely $U = \overline{u}/c$, $Z = z/H$ and $T = t F/(c h)$, we search for solutions $U \propto \widehat{U}(Z) \exp(\sigma T)$ that satisfy
\begin{equation}\label{eq:2o_eig}
\sigma \widehat{U} = L \widehat{U} \qquad \text{where} \qquad L \widehat{U} = \frac{\widehat{h}^2}{Re} \pdv[2]{z} \widehat{U} + 4 \mathrm{e}^{-Z/\widehat{h}} \qty(\widehat{U} - \frac{1}{\widehat{h}} \int_0^Z \widehat{U} \, \dd Z') \,.
\end{equation}
By defining $\widehat{Q} = \partial_Z \widehat{U}$ and $\widehat{U} = \partial_Z \widehat{F}$, Eq. \eqref{eq:2o_eig} can be written as a set of three first-order homogeneous equations:
\begin{equation}\label{eq:1o_homogen_eig}
\mqty(
\pdv{Z} & \frac{Re}{\widehat{h}^2} \qty(4 \mathrm{e}^{-Z/\widehat{h}} - \sigma) & - \frac{4 Re \mathrm{e}^{-Z/\widehat{h}}}{\widehat{h}^3} \\
1 & \pdv{Z} & 0 \\
0 & 1 & \pdv{Z} \\
)
\mqty(\widehat{Q} \\ \widehat{U} \\ \widehat{F}) =
\mqty(0 \\ 0 \\ 0) \,.
\end{equation}
The no-slip and stress-free boundary conditions are imposed at the bottom and top, respectively
\begin{equation}
\widehat{F}(0) = 0 \,, \quad \widehat{U}(0) = 0 \,, \quad \widehat{Q}(1) = 0 \,.
\end{equation}
The eigenspectrum of the matrix operator on the LHS of Eq. \eqref{eq:1o_homogen_eig} is solved numerically using Dedalus along with the Eigentools library (see Burns et al. (2020); Oishi et al. (2021)). The solutions are computed on a Chebyshev polynomial basis with $nz = 128$ grid points. Note that the operator $L$ is not self-adjoint, therefore the eigenspectrum cannot be directly considered as a natural basis to linearly decompose any solutions of the quasilinear model. Nevertheless, the properties of the eigenspectrum are helpful to understand the appearance of various bifurcations and each type of oscillation resolved by the quasilinear model.

\begin{figure}[h!]
\centering
\includegraphics[width=39pc]{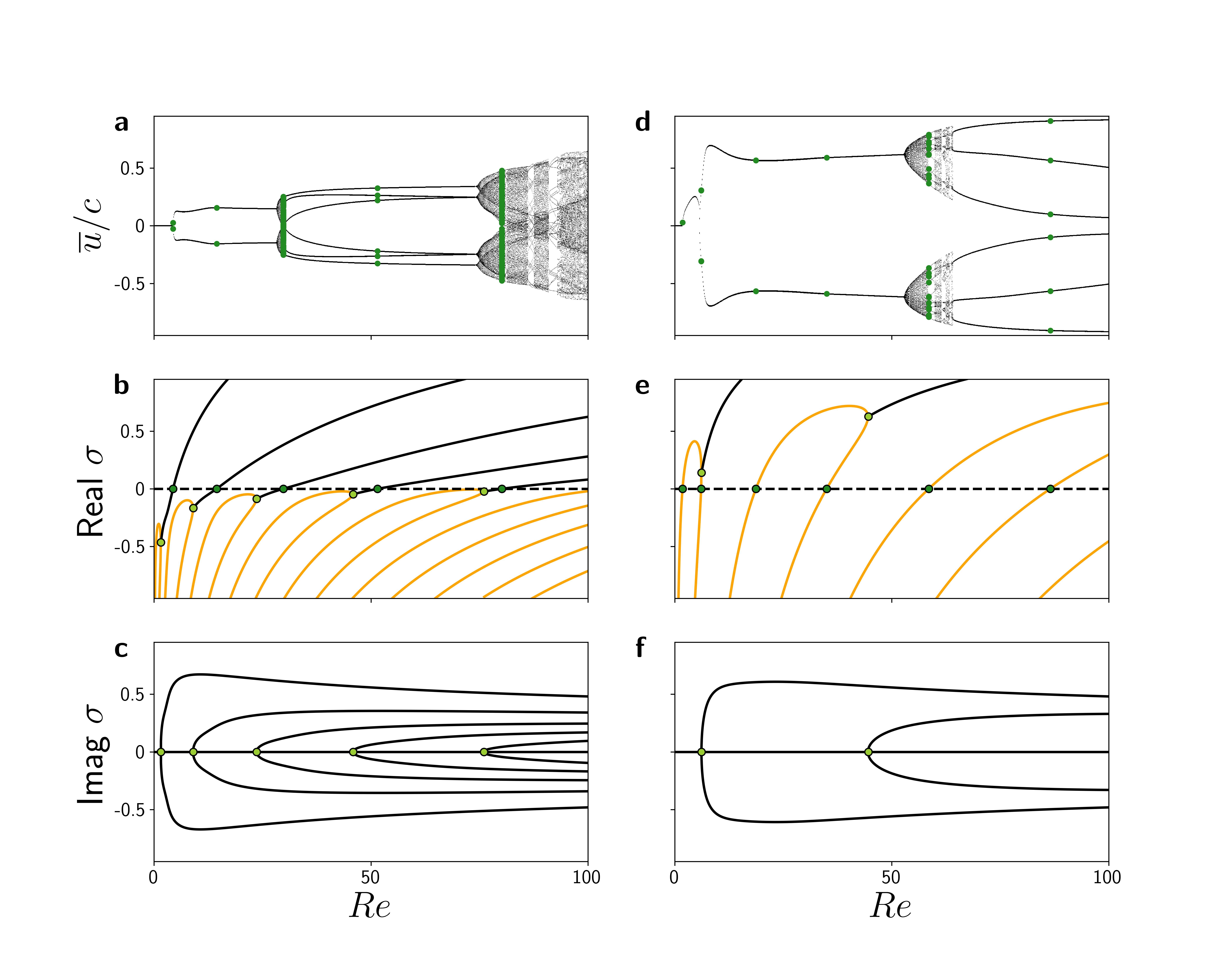}
\caption{Comparison of quasilinear simulations with $\widehat{h} = 1/3.5$ (left column) and $\widehat{h} = 1/1.25$ (right column), to their linearized version. (a), (d) Poincaré bifurcation diagram of quasilinear simulations. (b), (e) the real and (c), (f) the imaginary part of the eigenvalue $\sigma$.}
\label{fig:SM1_eigprob}
\end{figure}

Figure S\ref{fig:SM1_eigprob} compares bifurcation diagrams with solutions of the eigenvalue problem for two different values of $\widehat{h}$. At small $\widehat{h}$, the first unstable eigenmode is oscillatory, with a non-zero imaginary part (panels a-b-c). This corresponds to a supercritical Hopf bifurcation for the rest state in the quasi-linear system. At large $\widehat{h}$, the unstable eigenmodes are purely real (panels d-e-f). The first bifurcation corresponds to a pitchfork bifurcation leading to nontrivial stationary states, breaking the east-west symmetry in the quasilinear system. QBO-like states emerge in that case from secondary instabilities of those non-trivial stationary states, which cannot be detected from the operator obtained by linearizing around a state of rest. There is an intermediate case $\widehat{h} \in [\widehat{h}_c, \widehat{h}_{c2}]$, where the first and second unstable eigenmodes are real, while the third unstable mode is oscillatory. This corresponds to a subcritical Hopf bifurcation for the base state at rest. The critical value $\widehat{h} = \widehat{h}_c$ thus corresponds to a transition from a supercritical to a subcritical Hopf bifurcation.

\begin{figure}[h!]
\centering
\includegraphics[width=39pc]{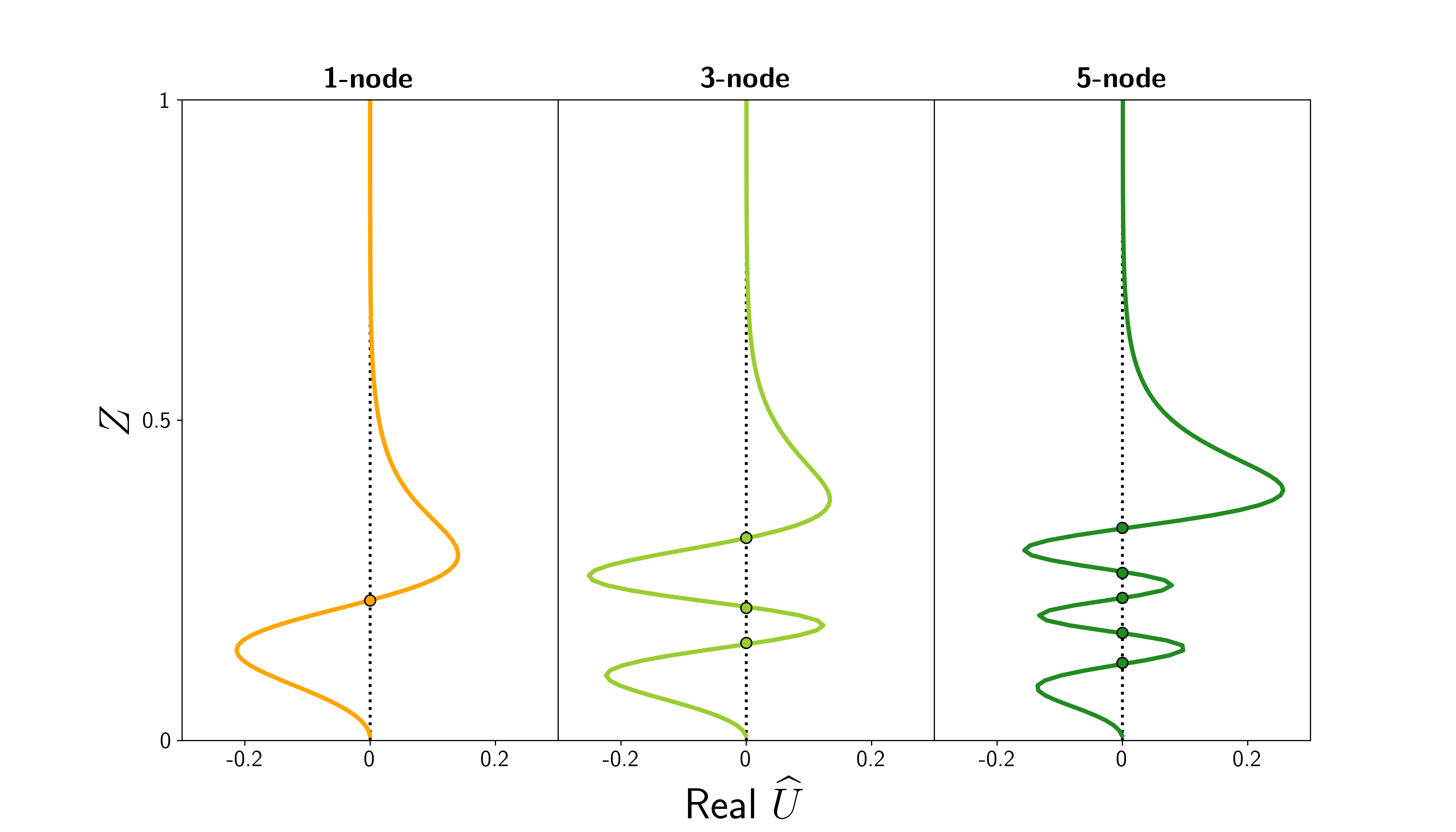}
\caption{Indexation of eigenmodes by the number of nodes (colored dots). Real part of $\widehat{U}$ for oscillatory modes indexed 1${}^\star$, 3${}^\star$ and 5${}^\star$ presenting 1, 3 and 5 nodes, respectively. The same indexation applies for stationary modes.}
\label{fig:SM2_eigv}
\end{figure}

The numbering of each vertical mode, shown in Fig. S\ref{fig:SM2_eigv}, is based on the maximum number of zeros in the column, excluding boundary conditions. For oscillatory modes, we consider the maximum through a complete oscillation.

\section*{\centering II. Periodicity recovery in the QBO${}_{(+)}$ region }
\begin{figure}[h!]
\centering
\includegraphics[width=30pc]{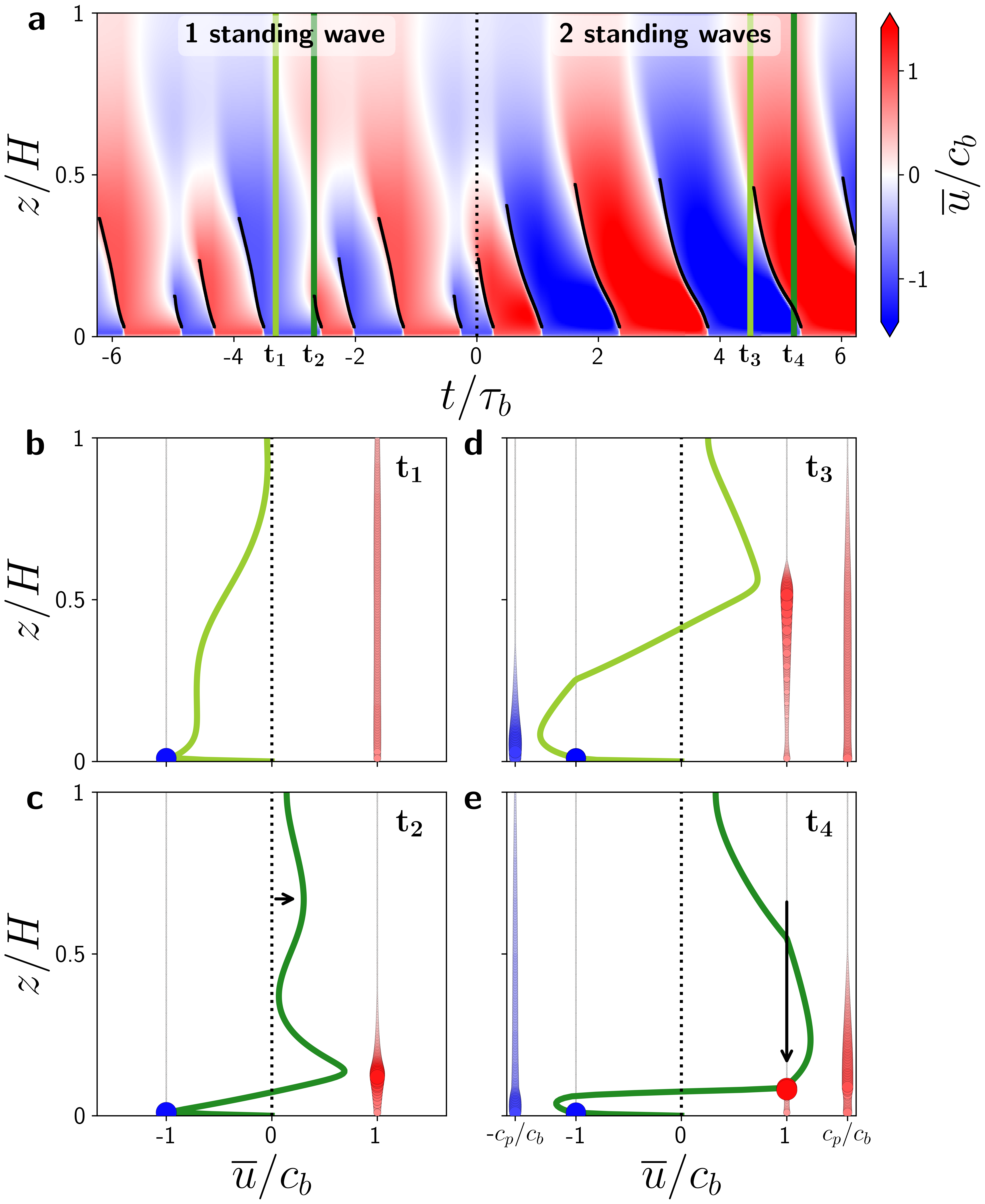}
\caption{Same as Figure 4, but for the case $c_p > c_b$. Black contours indicate critical layers of the background wave.}
\label{fig:SM3_LargeCL}
\end{figure}

In this section, we show the periodicity recovery in the region QBO${}_{(+)}$ ($c_p > c_b$) using an experiment similar to that shown in Fig. 4 (see Fig. S\ref{fig:SM3_LargeCL}). For this case, we selected $c_p/c_b = 4.4$ and $F_p/F_b = 0.8$. A synchronization mechanism analogous what's observed in the QBO${}_{(-)}$ occurs between critical layers of the background and the perturbation waves albeit with their role reversed. Here, the perturbation enhances the magnitude of the westward winds at $t_3$, leading to an injection of momentum by the background wave at higher altitudes. The rapid descent of the background wave's critical layer prevents the excitation of higher vertical modes, responsible for aperiodicity. A major difference with the periodicity recovery in the region QBO${}_{(-)}$ is that the perturbation doesn't entirely form a critical layer in this case. This suggests that critical layers of the fastest waves are not a necessary ingredient of this synchronization mechanism.

\section*{References}
Burns, K. J., G. M. Vasil, J. S. Oishi, D. Lecoanet, and B. P. Brown, 2020: Dedalus: A flexible framework for numerical simulations with spectral methods. \textit{Physical Review Research}, \textbf{2 (2)}, 023 068. \\
Oishi, J. S., K. J. Burns, S. E. Clark, E. H. Anders, B. P. Brown, G. M. Vasil, and D. Lecoanet, 2021: eigentools: A Python package for studying differential eigenvalue problems with an emphasis on robustness. \textit{Journal of Open Source Software}, \textbf{6 (62)}, 3079.

\end{document}